\documentclass[a4paper]{jpconf}
\usepackage{graphicx} \newcommand{\be}{\begin{equation}}\newcommand{\ee}{\end{equation}}
\newcommand{\bea}{\begin{eqnarray}}\newcommand{\eea}{\end{eqnarray}}
\newcommand{\beaa}{\begin{eqnarray}}\newcommand{\eeaa}{\end{eqnarray}}
\newcommand{\ba}{\begin{array}}\newcommand{\ea}{\end{array}}
\newcommand{\bit}{\begin{itemize}}\newcommand{\eit}{\end{itemize}}
\newcommand{\ben}{\begin{enumerate}}\newcommand{\een}{\end{enumerate}}

\def\pr{{\it Phys. Rev.}\ }

\def\pl{{\it Phys. Lett.}\ }

\def\rmp{{\it Rev. Mod. Phys.}\ }

\def\pr{{\it Phys. Rev.}\ }
\def\pl{{\it Phys. Lett.}\ }

\def\rmp{{\it Rev. Mod. Phys.}\ }

\def\lab{\label}
\def\lan{\langle}
\def\lf{\left}

\def\non{\nonumber}
\def\ran{\rangle}

\def\ri{\right}

\def\al{\alpha}\def\bt{\beta}

\def\te{\theta}

\def\si{\sigma}

\def\1{{_{1}}}\def\2{{_{2}}}

\begin{document}

\title{Particle mixing as possible explanation of the dark energy conundrum}

\author{Antonio Capolupo and Giuseppe Vitiello}

%\vspace{2mm}

\address{ Dipartimento di Matematica e Informatica,
 Universit\`a di Salerno and Istituto Nazionale di Fisica Nucleare,
 Gruppo Collegato di Salerno, 84100 Salerno, Italy.}

%\maketitle

\date{\today}

\vspace{2mm}

\begin{abstract}

The vacuum condensate due to neutrino and quark mixing behaves as a perfect fluid
and, at the present epoch, as a cosmological constant.
The very small breaking of the Lorentz invariance constrains today the value of
the dark energy.

\end{abstract}

%\pacs{98.80.Cq, 98.80. Hw, 04.20.Jb, 04.50+h}

%\maketitle

\section{Introduction}

Recent astrophysical data \cite{SCP}-\cite{WMAP-Five} represent observational
evidence of the fact that the universe today observed is flat and it is undergoing an
accelerated expansion.
The hypothesis is that such an expansion is driven by a fluid
called {\it dark energy} with negative
pressure, nearly homogeneously distributed and making up to $\sim
70\%$ of the energy content of the universe.

Even if it seems clear how dark energy works, its nature
remains an unsolved problem.
Therefore, the understanding of the dark energy has become one of the
main issue of modern physics. Its determination should provide
the  gravity vacuum state \cite{weinberg}, should clarify
 the mechanisms which led from the early universe to the today
observed large scale structures \cite{guth,linde}, and to
 represent the theoretical
solution to the observational data.

A lot of models have been proposed 
in order to explain the dark energy \cite{LCDMrev}-\cite{curvature}.
However, no comprehensive scheme is
available, which may fit the observations and frame them into a
 fundamental theory.

In this paper we report recent results according to which
the particle mixing phenomenon could explain the dark energy component of
the universe \cite{Capolupo:2008rz}. Responsible of such a component is
the non-perturbative vacuum structure associated with the particle mixing
\cite{Capolupo:2006et}-\cite{Blasone:2004yh}.

The estimated value of the cosmological constant is
imposed by the small breaking of the Lorentz invariance of the flavor vacuum
at the present epoch  \cite{Capolupo:2008rz}.

The paper is organized as follows. In Section II
we present the QFT formalism for mixed fields
\cite{BV95}-\cite{Blasone:2005ae}
 (for a detailed review see \cite{Capolupo:2004av}). 
We  introduce the particle mixing vacuum condensate
and we show that it is homogeneous and isotropic \cite{Capolupo:2008rz}. 
In Section III we exhibit the mixing contributions to the dark energy at the present epoch and
 Section IV is devoted to conclusions.

\section{Particle mixing vacuum condensate}

The mixing transformations among three
generations of Dirac fields are:
$ \Psi_f(x) \, = {\cal U} \, \Psi_m
(x)$,
where ${\cal U}$ is the CKM matrix
and $\Psi_m^T=(\psi_1,\psi_2,\psi_3)$ are
the fields with definite masses $m_{1} \neq m_{2} \neq m_{3}$.
The mixing relations can be written as
$\psi_{\si}^{\al}(x)\equiv G^{-1}_{\bf \te}(t) \,
\psi_{i}^{\al}(x)\, G_{\bf \te}(t), $ where $(\si,i)=(A,1), (B,2),
(C,3)$; $A$, $B$, $C$ denote lepton $(e, \mu, \tau)$
 or flavor $(d, s, b)$ indices. $G_{\bf \te}(t) $ is the mixing generator \cite{Capolupo:2008rz,yBCV02,Capolupo:2004av}.
The flavor annihilators, at each time,  are defined as:
$
\alpha _{{\bf k},\sigma}^{r}(t) \equiv G^{-1}_{\bf \te}(t)\;\alpha
_{{\bf k},i}^{r}(t)\;G_{\bf \te}(t)\,,$ and
$\beta _{{\bf k},\sigma}^{r}(t) \equiv G^{-1}_{\bf \te}(t)\;\beta
_{{\bf k},i}^{r}(t)\;G_{\bf \te}(t)\,.
$
They annihilate the flavor vacuum $ |0(t) \rangle_{f} = G^{-1}_{\bf \te}(t)\;
|0 \rangle_{m}\,, $
where $|0\rangle_{m}$ is the vacuum annihilated by $\alpha ^{r}_{{\bf k},i}$ and $ \beta ^{r
}_{{\bf k},i}$, $ i=1,2,3 \;, \;r=1,2$.
In the infinite volume limit  $|0(t) \rangle_{f}$ turns out to be unitarily
inequivalent to the vacuum  $|0
\rangle_{m}$ \cite{BV95}, \cite{hannabuss}.
Moreover, $|0(t) \rangle_{f}$ is a coherent condensate of particles.
In the reference frame such that
${\bf k}=(0,0,|{\bf k}|)$,  the numbers of condensed particles are:
\bea \lab{V1}
{\cal N}^{\bf k}_1\, = \,_{f}\langle0(t)|N^{{\bf
k},r}_{\al_{1}} |0(t)\ran_{f} \, = \,_{f}\langle0(t)|N^{{\bf
k},r}_{\bt_{1}}|0(t)\ran_{f}\, = \, s^{2}_{12}c^{2}_{13}\,|V^{{\bf
k}}_{12}|^{2}+ s^{2}_{13}\,|V^{{\bf k}}_{13}|^{2}\,,
\eea
and similar relations for ${\cal N}^{\bf k}_2$, ${\cal N}^{\bf k}_3$.
In Eq.(\ref{V1}), $N^{{\bf k},r}_{\al_{i}} \,=\, \alpha_{{\bf k},i}^{r \dag} \alpha_{{\bf k},i}^{r} \,$,
 $N^{{\bf k},r}_{\beta_{i}} \,=\, \beta_{{\bf k},i}^{r \dag} \beta_{{\bf k},i}^{r} \,$
with $i =1,2,3,$ and
 $V^{{\bf k}}_{ij}$ are the Bogoliubov coefficients entering the mixing transformations
(see  Refs.\cite{Capolupo:2008rz,yBCV02,Capolupo:2004av}).

As shown in Ref.\cite{Capolupo:2008rz}, the vacuum condensate
 due to neutrino and quark mixing behaves as a perfect fluid.
Indeed considering the Minkowski metric,  the energy-momentum tensor density
 ${\cal T}_{\mu\nu}(x) $ for the fermion fields $\psi_i$,  $i=1,2,3$ is \cite{Itz}
$
 :{\cal T}_{\mu\nu}(x):\, = \,  \frac{i}{2}:\left({\bar \Psi}_{m}(x)\gamma_{\mu}
\stackrel{\leftrightarrow}{\partial}_{\nu} \Psi_{m}(x)\right): \,$.
Then the energy momentum tensor density of the vacuum condensate is given by
\bea \label{T-Cond}
{\cal T}_{\mu\nu}^{cond}(x)={}_{f}\lan 0(t) |:{\cal T}_{\mu\nu}(x):| 0(t)\ran_{f}\,,
\eea
for which the off-diagonal components  are zero:
\bea
\int d^{3}x \;{}_{f}\lan 0(t) |{\cal T}_{0j}(x)| 0(t)\ran_{f}\,
\, = \,
\int d^{3}x \;{}_{f}\lan 0(t) |{\cal T}_{j\,l}(x)| 0(t)\ran_{f}\,
\,=\, 0\,,
\eea
where  $j \neq 0\,,$ and $j \neq l$, respectively. Thus ${\cal T}_{\mu\nu}^{cond}$
is homogeneous and isotropic and can be written as
\bea\label{Tmu,nu}
{\cal T}_{\mu\nu}^{cond}\,
 = \, diag
({\cal T}_{00}^{cond}\,,{\cal T}_{11}^{cond}\,,{\cal T}_{22}^{cond}\,
,{\cal T}_{33}^{cond}\,)\,,
\eea
where
\bea\label{A27}
\int d^{3}x \;{\cal T}_{00}^{cond}(x)
& = & \int d^{3}x \;{}_{f}\lan 0(t) |:{\cal T}_{00}(x):| 0(t)\ran_{f}\,
\,=\, 4 \sum_{i}  \int d^{3}{\bf k} \,
\omega_{k,i}\;  {\cal N}^{\bf k}_i\,,
\\\label{A28}
\int d^{3}x \;{\cal T}_{jj}^{cond}(x)
& = & \int d^{3}x \;{}_{f}\lan 0(t) |:{\cal T}_{jj}(x):| 0(t)\ran_{f}\,
\,=\, 4 \sum_{i}\int {d^{3}{\bf k}} \, \frac{k_j k_j}{\;\omega_{k,i}}\;
 {\cal N}^{\bf k}_i\,,
\eea
(no summation on $j$ is intended) and ${\cal N}^{\bf k}_i$  are the numbers of particles condensed in
the vacuum (see Eq.(\ref{V1})).

\section{Particle mixing and dark energy}

In Ref.\cite{Capolupo:2008rz} it has been shown that the energy density due to the vacuum
condensate arising from particle mixing gives a contribution to the vacuum
energy which evolves dynamically (see also Refs.\cite{Capolupo:2006et}, \cite{Capolupo:2007hy}).
Moreover, it has been remarked that the negligible
breaking of the Lorentz invariance of the vacuum is responsible of the very small value of the
dark energy at the present epoch \cite{Capolupo:2008rz}.
In this Section we report the main results obtained in Ref.\cite{Capolupo:2008rz}.

Let us write the energy-momentum tensor density
 ${\cal T}_{\mu\nu}(x) $ for the fermion fields $\psi_i$,  $i=1,2,3$  as \cite{leite}
\bea\non
 :{\cal T}_{\mu\nu}(x): & = &
 : \Sigma_{\mu\nu}(x):\,+\,:{\cal V}_{\mu\nu}(x):
 \\\non
 & = &
 : \Big\{\frac{i}{2}\left({\bar \Psi}_{m}(x)\gamma_{\mu}
\stackrel{\leftrightarrow}{\partial}_{\nu} \Psi_{m}(x)\right)
- \eta_{\mu\nu} \lf[\frac{i}{2} {\bar \Psi}_{m}(x)  \gamma^{\alpha}
\stackrel{\leftrightarrow}{\partial}_{\alpha}  \Psi_{m}(x) \ri]
\\
& + & \eta_{\mu\nu}  \lf[ {\bar \Psi}_{m}(x) \, \textsf{M}_d \,  \Psi_{m}(x) \ri]\Big\}: \;,
\eea
where
\bea
:\Sigma_{\mu\nu}(x): & = &
:\lf\{\frac{i}{2}\left({\bar \Psi}_{m}(x)\gamma_{\mu}
\stackrel{\leftrightarrow}{\partial}_{\nu} \Psi_{m}(x)\right)
- \eta_{\mu\nu} \lf[\frac{i}{2} {\bar \Psi}_{m}(x)  \gamma^{\alpha}
\stackrel{\leftrightarrow}{\partial}_{\alpha}  \Psi_{m}(x) \ri]\ri\}: \;,
\\
:{\cal V}_{\mu\nu}(x): & = &
 \eta_{\mu\nu} : \lf[ {\bar \Psi}_{m}(x) \, \textsf{M}_d \,  \Psi_{m}(x) \ri]: \;,
 \eea
 $\textsf{M}_d= diag(m_{1}, m_{2}, m_{3})$, $\Psi_{m} = (\psi_1, \psi_2, \psi_3)^{T}$
 and $\eta_{\mu\nu} = diag (1,-1,-1,-1)$.

In any epochs the contributions of the
particle mixing to the vacuum energy density $\rho_{mix}$
and to the vacuum pressure $ p_{mix}$ are given respectively by
the $(0,0)$ and the $(j,j)$ components of the energy momentum tensor density of the vacuum condensate:
 \bea\label{ro}
\rho_{mix} & \equiv & \frac{1}{ V}\; \eta^{00}\; \int d^{3}x \;{\cal T}_{00}^{cond}(x)  ~,
 \\
 \label{p-mix}
p_{mix} & \equiv &  -\frac{1}{ V}\; \eta^{jj} \; \int d^{3}x \;{\cal T}_{jj}^{cond}(x)  ~,
 \eea
  where no summation on the index $j$ is intended. By using Eqs.(\ref{A27}) and (\ref{A28})
  we have:
\bea\label{rho-past}
\rho_{mix}
& = & \frac{2}{\pi} \sum_{i}\,  \int d k \, k^{2}\,
\omega_{k,i}\;  {\cal N}^{\bf k}_i\,,
\\\label{p-past}
p_{mix}
& = &  \frac{2}{3 \pi} \sum_{i}\,\int d k \, k^{2} \, \frac{k^{2}}{\;\omega_{k,i}}\;
 {\cal N}^{\bf k}_i\,.
\eea

We note that the use of the identity
$\omega_{k,i} = \frac{k^{2}}{\omega_{k,i}} + \frac{m_{i}^{2}}{\omega_{k,i}}$
enables to write $\rho_{mix}$ as
\bea\label{energy(delta)} \rho_{mix}= \Sigma_{mix}+
{\cal V}_{mix} \eea
where the kinetic term $\Sigma_{mix}$ and the potential term
${\cal V}_{mix}$ are given by
\bea \label{delta2}
\Sigma_{mix} \, = \, \frac{2}{\pi} \sum_{i}  \int_{0}^{K} dk \, k^{2}\;
\frac{k^{2}}{\omega_{k,i}}\;{\cal N}^{\bf k}_i\,, \qquad
 {\cal V}_{mix} \, = \, \frac{2}{\pi} \sum_{i}  \int_{0}^{K} dk \, k^{2}\;
\frac{m_{i}^{2}}{\omega_{k,i}}\;{\cal N}^{\bf k}_i\,.
 \eea

However, in the present epoch, the very small breaking of the Lorentz invariance
\cite{WMAP-Five}, imposes that
the vacuum expectation values of ${\cal T}_{\mu\nu}(x)$
are space-time independent.
This implies that the kinematical part $\Sigma^{cond}_{\mu\nu}$
of ${\cal T}^{cond}_{\mu\nu}$ is today very small \cite{Capolupo:2008rz}:
\bea\label{lorentzInv}
\Sigma^{cond}_{\mu\nu}\,=\;{}_{f}\lan
0(t) |:\Sigma_{\mu\nu}(x):| 0(t)\ran_{f}\,\simeq 0
\eea
and ${\cal T}_{\mu\nu}^{cond}$ is given by:
\bea\label{Tcond} {\cal T}_{\mu\nu}^{cond} \simeq \;{}_{f}\lan
0(t) |:{\cal V}_{\mu\nu}(x):| 0(t)\ran_{f}\, =\,
\eta_{\mu\nu}\;{}_{f}\lan
0(t) |:{\bar \Psi}_{m}(x)\;\textsf{M}_d\;\Psi_{m}(x):| 0(t)\ran_{f}\,.
 \eea

By using  Eqs.(\ref{Tmu,nu}) and (\ref{Tcond}), at the present epoch, we obtain:
\bea\label{Tcond1}\non
\frac{1}{V}\;  \int d^{3}x \;{\cal T}_{\mu\nu}^{cond}(x) &=& diag
(\rho_{mix}\,,p_{mix}\,,p_{mix}\,,p_{mix})
\\
&=& \eta_{\mu\nu}\;\sum_{i}m_{i}\int \frac{d^{3}x}{(2\pi)^3}\;{}_{f}\lan
0 |:\bar{\psi }_{i}(x)\psi_{i}(x):| 0\ran_{f}\,.
 \eea
The today state equation  is then given by:
$\rho_{mix} \simeq -p_{mix}$, that is,
 the adiabatic index  is
\bea
w_{mix}\; = \; p_{mix}/ \rho_{mix}\; \simeq \; - 1\;.
\eea

This means that the vacuum condensate coming from  particle
mixing, today, mimics the behavior of the cosmological constant
\cite{Capolupo:2006et}, \cite{Capolupo:2007hy}. From Eq.(\ref{Tcond1}),
introducing the cut-off on the momenta $K$, we derive  $\rho_{mix}$
 \cite{Capolupo:2008rz}:
 \bea \label{cost1}
\rho_{mix} & \simeq & \frac{2}{\pi} \sum_{i} \int_{0}^{K} dk \, k^{2}\;
\frac{m_{i}^{2}}{\omega_{k,i}} \;{\cal N}^{\bf k}_i\,.
 \eea

The integral (\ref{cost1}) diverges in $K$ as $m_{i}^{4}\,\log\lf( 2K /m_{j}\ri)$,
 with $i,j = 1,2,3$ \cite{Capolupo:2007hy}.
However, the value close to $-1$ of $w_{mix}$
 at the present epoch constrains the value of $K$ and consequently the
 value of $\rho_{mix}$ \cite{Capolupo:2008rz}.
Indeed, by comparing Eqs.(\ref{rho-past}) and (\ref{cost1}),
we note that the condition (\ref{lorentzInv}) for the $(0,0)$
component of ${\cal T}_{\mu\nu}$ imposes
 $\Sigma_{mix} \ll \, {\cal V}_{mix}$, where  $\Sigma_{mix}$ and ${\cal V}_{mix}$
 are given in Eq.(\ref{delta2}).
Such a condition is satisfied when $K$ in Eq.(\ref{delta2}) is
$ K \ll  \sqrt[3]{m_{1} m_{2} m_{3}}\,.$
For this reason here
 we do not need to consider the regularization problem of the
ultraviolet divergence of the integral (\ref{cost1}).

We now show that particular values of $w_{mix}$, for neutrinos and for quarks,
lead to contributions of the dark energy that are in agreement with its estimated upped bound.
To do that we derive an expression of $w_{mix} $ as function of the cut-off $K$ \cite{Capolupo:2008rz}.

Let us consider  the adiabatic expansion of a sphere of volume $V$.
Since the pressure $p$, at which the sphere expands, does work,
the total energy, $E = \rho \,V$, varies:
$dE = -p\, d V $. That is $ \rho\, d V +
 V\, d\rho= -p\, d V $,
that can be expressed as
$
d[(\rho + p) V] =0\,,
$
and then
\bea\label{energy-Pcost1}
\rho + p  = \frac{const}{V}\,.
\eea

Thus if the volume is very large  ($V \rightarrow \infty $),  we have
$\rho \simeq - p $ and $w =p/\rho \simeq -1$.

Being $\rho = \Sigma + {\cal V}$
where $\Sigma$ and ${\cal V}$ are the kinetic and the potential terms respectively,
and taking into account Eq.(\ref{energy-Pcost1}), for a volume $V \rightarrow \infty $,
we obtain  $ \rho = \Sigma + {\cal V} \simeq -p$.
Taking into account the condition $\Sigma \ll \, {\cal V}$
imposed by the negligible breaking of the Lorentz invariance
at the present epoch, we have $\rho \simeq {\cal V} \simeq - p$.
Then, from the equality:
 $\rho = \frac{const}{V} - p = \Sigma + {\cal V}$,  we have
 $ \Sigma \simeq \frac{const}{V} \simeq 0, $ for $V \rightarrow \infty$.
This implies that, at the present epoch, $\rho$ can be written as
 $\rho = \Sigma - p \simeq -p$. Then, in the case of the flavor vacuum condensate,
 $w_{mix}$ is given by:
\bea\label{Wmix} w_{mix}  =
\frac{p_{mix} }{\Sigma_{mix}  - p_{mix} }\,.
\eea

Since today $\Sigma_{mix} \sim 0$, then $w_{mix} \sim -1$.
Eq.(\ref{Wmix}) gives an expression of
$w_{mix} $ as function of $K$ (since $\Sigma_{mix}$ and $p_{mix}$ are function of $K$).

By using different values of $w_{mix}$ close to  $-1$, we compute the contributions given to the dark energy by the particle
mixing condensates, at the present epoch. We found the following results \cite{Capolupo:2008rz}:

\vspace{3mm}
\textit{Neutrino mixing condensate contribution:}
\vspace{2mm}

$\rho^{\nu}_{mix} \sim  10^{-47} GeV^{4}$ for $-0.98\leq w^{\nu}_{mix} \leq -0.97\,.$
Such contributions are compatible with the estimated
upper bound of the dark energy and $w^{\nu}_{mix}$ is in agreement with
the constraint on the dark energy state equation \cite{WMAP-Five}.

For $w^{\nu}_{mix} < -0.98$ we have negligible contributions
of $\rho^{\nu}_{mix}$. The results we found are
dependent on the neutrino mass values one uses.
(We have considered values of the neutrino masses
such that the experimental values of squared mass difference \cite{Altarelli:2007gb}
 are satisfied, as for example:
 $m_{1} = 4.6 \times 10^{-3}eV$,
$m_{2} = 1 \times 10^{-2}eV$,  $m_{3} = 5 \times 10^{-2}eV$).

\vspace{3mm}
\textit{Quark mixing condensate contribution:}
\vspace{2mm}

A contribution compatible with the estimated
upper bound of the dark energy: $\rho^{q}_{mix} = 1.5 \times 10^{-47} GeV^{4}$
is found for $w^{q}_{mix} =-1$ \cite{Capolupo:2008rz}.
Very small deviations from the value $w^{q}_{mix} = -1$ 
give rise to contributions of $\rho^{q}_{mix}$
that are beyond the accepted upper bound of the dark energy.

\section{Conclusions and discussion}

In this report we have shown that the vacuum condensate from particle mixing 
provides a contribution to the dark energy which is compatible with the
estimated value of the cosmological constant. Such  value  is imposed by the small
 breaking of the Lorenz of the flavor vacuum at the present epoch.

\section*{Acknowledgements}

Support from INFN and Miur is acknowledged.

\medskip

\section*{References}

\bibliography{apssamp}% Produces the bibliography via BibTeX.

\begin{thebibliography}{99}

%%\cite{SNO}
%\bibitem{SNO}
%%"Direct Evidence for Neutrino Flavor Tranformation from Neutral-Current Interactions in the Sudbury Neutrino Observatory"
%SNO Collaboration, Phys.\ Rev.\ Lett. {\bf 89}, No. 1, 011301 (2002).
%
%%\cite{K2K}
%\bibitem{K2K}
% %"Evidence for muon neutrino oscillation in an accelerator-based expetiment"
%K2K collaboration, E. Aliu et al, Phys. \ Rev. \ Lett. {\bf 94}, 081802
%(2005).

\bibitem{SCP}
S. Perlmutter et al., ApJ {\bf 517}, 565 (1999);
\\
R.A. Knop et al.,
ApJ {\bf 598}, 102 (2003).

\bibitem{HZT}
A.G. Riess et al., AJ {\bf 116}, 1009 (1998);
\\
J.L. Tonry et al., ApJ
{\bf 594}, 1 (2003).

\bibitem{Boomerang}
P. de Bernardis et al., Nature {\bf 404}, 955 (2000).

\bibitem{Maxima}
R. Stompor et al., ApJ  {\bf 561}, L7 (2001).

\bibitem{WMAP}
D.N. Spergel et al. ApJS {\bf 148}, 175 (2003).

\bibitem{Riess04}
A.G. Riess et al., ApJ {\bf 607}, 665 (2004).

\bibitem{WMAP-Five}
G.~Hinshaw, et al.,
%Five-Year Wilkinson Microwave Anisotropy Probe (WMAP) Observations: Data Processing, Sky Maps, and Basic Results
astro-ph/0803.0732v1;
%
\\
R.~Hill, et al.,
%Five-Year Wilkinson Microwave Anisotropy Probe (WMAP) Observations: Beam Maps and Window Functions
astro-ph/0803.0570v1;
%
\\
B.~Gold, et al.,
%Five-Year Wilkinson Microwave Anisotropy Probe (WMAP) Observations: Galactic Foreground Emission
astro-ph/0803.0715v1;
%
\\
E.~Wright, et al.,
%The Wilkinson Microwave Anisotropy Probe (WMAP) Source Catalog
astro-ph/0803.0577v1;
%
\\
M.~Nolta, et al.,
%Five-Year Wilkinson Microwave Anisotropy Probe (WMAP) Observations: Angular Power Spectra
astro-ph/0803.0593v1;
%
\\
J.~Dunkley, et al.,
%Five-Year Wilkinson Microwave Anisotropy Probe (WMAP) Observations: Likelihoods and Parameters from WMAP Data
astro-ph/0803.0586v1;
%
\\
E.~Komatsu, et al.,
%Five-Year Wilkinson Microwave Anisotropy Probe (WMAP) Observations: Cosmological Interpretation
astro-ph/0803.0547v1.
%


\bibitem{weinberg}
S. Weinberg, \rmp {\bf 61},1  (1989).

\bibitem{guth}
A. Guth, \pr  D {\bf 23}, 347 (1981); 
\\
A. Guth,  \pl B {\bf 108 }, 389
(1982).

\bibitem{linde}
A.D. Linde, \pl B {\bf  108}, 389 (1982); 
\\
A.D. Linde, \pl B {\bf
114}, 431 (1982); 
\\
A.D. Linde, \pl  B {\bf 129}, 177 (1983); 
\\
A.D. Linde  \pl  B {\bf 238}, 160 (1990).


\bibitem{LCDMrev}
 V. Sahni, A. Starobinski, Int. J. Mod. Phys. D {\bf 9}, 373 (2000).


%\cite{Sahni:2004ai}
\bibitem{Sahni:2004ai}
  V.~Sahni,
  %``Dark matter and dark energy,''
  Lect.\ Notes Phys.\  {\bf 653}, 141 (2004);
  %%CITATION = LNPHA,653,141;%%
%
\\
  S.M. Carroll,
%``The Cosmological constant,''
Living Rev.\ Rel. {\bf 4}, 1 (2001);
 %[arXiv:astro-ph/0004075].
  %%CITATION = ASTRO-PH 0004075;%%
%
\\
P.J.E. Peebles, B. Ratra,
%``The Cosmological constant and dark energy,''
Rev. Mod. Phys. {\bf 75}, (2003);
\\
%\cite{Copeland:2006wr}
%\bibitem{Copeland:2006wr}
  E.~J.~Copeland, M.~Sami and S.~Tsujikawa,
  %``Dynamics of dark energy,''
  Int.\ J.\ Mod.\ Phys.\  D {\bf 15}, 1753 (2006).
  %[arXiv:hep-th/0603057].
  %%CITATION = IMPAE,D15,1753;%%



\bibitem{QuintRev}
T. Padmanabhan, Phys. Rept. {\bf 380}, 235 (2003).


\bibitem{Chaplygin}
A. Kamenshchik, U. Moschella, V. Pasquier, Phys. Lett. B {\bf 511},
265 (2001).

\bibitem{tachyon}
 T. Padmanabhan, Phys. Rev. D {\bf 66}, 021301 (2002).

\bibitem{Bassett}
B.A. Bassett, M. Kunz, D. Parkinson, C. Ungarelli, Phys. Rev. D
{\bf 68}, 043504 (2003).


\bibitem{Cardassian}
K. Freese, M. Lewis, Phys. Lett. B {\bf 540}, 1 (2002).

\bibitem{DGP}
G.R. Dvali, G. Gabadadze, M. Porrati, Phys. Lett. B {\bf 485}, 208
(2000).

\bibitem{curvature}
 S.Nojiri and S.D. Odintsov, Phys. Rev. D {\bf 68}, 123512 (2003);
 %
 \\
S.M. Carroll,  V. Duvvuri, M. Trodden, M.S. Turner, Phys. Rev. D
{\bf 70}, 043528 (2004);
%
\\
 G. Allemandi, A. Borowiec, M.
 Francaviglia, Phys. Rev. D {\bf 70}, 103503 (2004).


%\cite{Capolupo:2008rz}
\bibitem{Capolupo:2008rz}
  A.~Capolupo, S.~Capozziello and G.~Vitiello,
  %``Dark energy and particle mixing,''
Phys. Lett. A  {\bf 373}, 601-610 (2009).
  %%CITATION = ARXIV:0809.0085;%%

%\cite{Capolupo:2006et}
\bibitem{Capolupo:2006et}
  A.~Capolupo, S.~Capozziello and G.~Vitiello,
  %``Dark energy explained by the mixing of neutrinos,''
  Phys. Lett. A  {\bf 363}, 53 (2007);
   %%CITATION = ASTRO-PH 0602467;%%
   %
\\
A.~Capolupo, S.~Capozziello and G.~Vitiello,
  %``Dark energy induced by neutrino mixing,''
  J.\ Phys.\ Conf.\ Ser.\  {\bf 67}, 012032 (2007).
  %[arXiv:hep-th/0612035].
  %%CITATION = 00462,67,012032;%%
%

%\cite{Capolupo:2007hy}
  \bibitem{Capolupo:2007hy}
  A.~Capolupo, S.~Capozziello and G.~Vitiello,
  %``Dark energy, cosmological constant and neutrino mixing,''
Int. J. Mod. Phys. A {\bf 23}, 4979-4990 (2008).
  %%CITATION = ARXIV:0705.0319;%%

%\cite{Blasone:2004yh}
\bibitem{Blasone:2004yh}
  M.~Blasone, A.~Capolupo, S.~Capozziello, S.~Carloni and G.~Vitiello,
  %``Neutrino mixing contribution to the cosmological constant,''
  Phys.\ Lett.\  A {\bf 323}, 182 (2004);
  %[arXiv:gr-qc/0402013].
  %%CITATION = PHLTA,A323,182;%%
\\
%\cite{Blasone:2007jm}
%\bibitem{Blasone:2007jm}
  M.~Blasone, A.~Capolupo, S.~Capozziello and G.~Vitiello,
  %``Neutrino mixing, flavor states and dark energy,''
  Nucl.\ Instrum.\ Meth.\  A {\bf 588}, 272 (2008).
 %[arXiv:0711.0939 [hep-th]].
  %%CITATION = NUIMA,A588,272;%%

\bibitem{BV95}
M. Blasone and G. Vitiello,
%``Quantum field theory of fermion mixing,''
Annals Phys.\ {\bf 244}, 283 (1995).
% [hep-ph/9501263].

%\cite{Fujii:1999xa}
\bibitem{Fujii:1999xa}
K.~Fujii, C.~Habe and T.~Yabuki, Phys.\ Rev.\ D {\bf 59}, 113003
(1999);
\\
K.~Fujii, C.~Habe and T.~Yabuki,
 Phys.\ Rev.\ D {\bf 64}, 013011 (2001).

\bibitem{JM01}
C.R. Ji, Y. Mishchenko, Phys. Rev. D {\bf 64}, 076004 (2001);
\\
C.R. Ji, Y. Mishchenko,
Phys. Rev. D {\bf 65}, 096015 (2002).


\bibitem{hannabuss}
K.~C.~Hannabuss and D.~C.~Latimer,
%``The quantum field theory of fermion mixing,''
J.\ Phys.\ A {\bf 33}, 1369 (2000);
%%CITATION = JPAGB,A33,1369;%
\\
K.~C.~Hannabuss and D.~C.~Latimer,
%``Fermion mixing in quasifree states,''
J.\ Phys.\ A {\bf 36}, L69 (2003).
%%CITATION = HEP-TH 0207268;%%

\bibitem{yBCV02}
M.~Blasone, A.~Capolupo and G.~Vitiello,
%``Quantum field theory of
%three flavor neutrino mixing and oscillations with CP violation,''
Phys.\ Rev.\ D {\bf 66}, 025033 (2002).
%%CITATION = HEP-TH 0204184;%%

%\bibitem{BCRV01}
%M.~Blasone, A.~Capolupo, O.~Romei and G.~Vitiello,
%%``Quantum field theory of boson mixing,''
%Phys.\ Rev.\ D {\bf 63}, 125015 (2001),
%%[hep-ph/0102048].
%%%CITATION = HEP-PH 0102048;%
%A.~Capolupo, C.~R.~Ji, Y.~Mishchenko and G.~Vitiello,
%%``Phenomenology of flavor oscillations with non-perturbative effects from
%%quantum field theory,''
%Phys.\ Lett.\ B {\bf 594}, 135 (2004). *********
%%[arXiv:hep-ph/0407166].
%%%CITATION = HEP-PH 0407166;%%

%\cite{Capolupo:2004av}
\bibitem{Capolupo:2004av}
A.~Capolupo, Ph.D. Thesis
%``Aspects of particle mixing in quantum field theory,''
%arXiv:
[hep-th/0408228].
%%CITATION = HEP-TH 0408228;%%

%\cite{Blasone:2001du}
\bibitem{Blasone:2001du}
  M.~Blasone, A.~Capolupo, O.~Romei and G.~Vitiello,
  %``Quantum field theory of boson mixing,''
  Phys.\ Rev.\  D {\bf 63}, 125015 (2001);
 %[arXiv:hep-ph/0102048].
  %%CITATION = PHRVA,D63,125015;%%
\\
%\cite{Capolupo:2004pt}
%\bibitem{Capolupo:2004pt}
  A.~Capolupo, C.~R.~Ji, Y.~Mishchenko and G.~Vitiello,
  %``Phenomenology of flavor oscillations with non-perturbative effects from
  %quantum field theory,''
  Phys.\ Lett.\  B {\bf 594}, 135 (2004).
 %[arXiv:hep-ph/0407166].
  %%CITATION = PHLTA,B594,135;%%

%\cite{Blasone:2005ae}
\bibitem{Blasone:2005ae}
M.~Blasone, A.~Capolupo, F.~Terranova and G.~Vitiello,
%``Lepton charge and neutrino mixing in decay processes,''
Phys.\ Rev.\ D {\bf 72}, 013003 (2005);
  %[arXiv:hep-ph/0505178].
  %%CITATION = HEP-PH 0505178;%%
  %
\\
%\cite{Blasone:2005tk}
%\bibitem{Blasone:2005tk}
  M.~Blasone, A.~Capolupo and G.~Vitiello,
  %``Mixing in quantum field theory,''
  Acta Phys.\ Polon.\  B {\bf 36}, 3245 (2005);
  %%CITATION = APPOA,B36,3245;%%
\\
%\cite{Blasone:2006jx}
%\bibitem{Blasone:2006jx}
  M.~Blasone, A.~Capolupo, C.~R.~Ji and G.~Vitiello,
  %``Flavor charges and flavor states of mixed neutrinos,''
 hep-ph/0611106.
  %%CITATION = HEP-PH/0611106;%%


\bibitem{Itz}
C.~Itzykson and J.~B.~Zuber, {\it Quantum Field Theory},
(McGraw-Hill, New York, 1980);
\\
S. Schweber, {\it An itroduction Relativistic Quantum Field Theory},
(Harper and Row, 1961).

\bibitem{leite}
J.~Leite ~Lopez, {\it Gauge Field Theories},
(Pergamon press, 1981).


%\cite{Altarelli:2007gb}
\bibitem{Altarelli:2007gb}
  G.~Altarelli,
  %``Lectures on Models of Neutrino Masses and Mixings,''
  hep-ph/0711.0161.
  %%CITATION = ARXIV:0711.0161;%%

%
%\bibitem{WMAP-three}
%D.N.~Spergel, et al.
%%Three-Year Wilkinson Microwave Anisotropy Probe (WMAP) Observations: Implications for Cosmology
%Astr. Jou. Suppl. Ser., {\bf 170}: 377-408, (2007).
%
%
%%\cite{PDG}
%\bibitem{PDG}
%W.-M. Yao et al. (Particle Data Group), J. Phys. G {\bf 33}, 1 (2006).



\end{thebibliography}

\end{document}